\documentclass[9pt,twocolumn,twoside]{opticajnl}
\journal{opticajournal} 

\setboolean{shortarticle}{true}

\usepackage{lineno}

\title{Highly stable power control for chip-based continuous-variable quantum key distribution system}

\author[1,2]{Yiming Bian}
\author[3,*]{Yang Li}
\author[1,2]{Xuesong Xu}
\author[3]{Tao Zhang}
\author[3]{Yan Pan}
\author[3]{Wei Huang}
\author[1,2]{Song Yu}
\author[1,4]{Lei Zhang}
\author[1,2,$\dagger$]{Yichen Zhang}
\author[3]{Bingjie Xu}

\affil[1]{State Key Laboratory of Information Photonics and Optical Communications, Beijing University of Posts and Telecommunications, Beijing, 100876, China}
\affil[2]{School of Electronic Engineering, Beijing University of Posts and Telecommunications, Beijing, 100876, China}
\affil[3]{Science and Technology on Communication Security Laboratory, Institute of Southwestern Communication, Chengdu 610041, China}
\affil[4]{School of Integrated Circuits, Beijing University of Posts and Telecommunications, Beijing, 100876, China}

\affil[*]{yishuihanly@pku.edu.cn}
\affil[$\dagger$]{zhangyc@bupt.edu.cn}

\begin{abstract}
Quantum key distribution allows secret key generation with information theoretical security. It can be realized with photonic integrated circuits to benefit the tiny footprints and the large-scale manufacturing capacity. Continuous-variable quantum key distribution is suitable for chip-based integration due to its compatibility with mature optical communication devices. However, the quantum signal power control compatible with the mature photonic integration process faces difficulties on stability, which limits the system performance and causes the overestimation of secret key rate that opens practical security loopholes. 
Here, a highly stable chip-based quantum signal power control scheme based on a biased Mach-Zehnder interferometer structure is proposed, theoretically analyzed and experimentally implemented with standard silicon photonic techniques. Simulations and experimental results show that the proposed scheme significantly improves the system stability, where the standard deviation of the secret key rate is suppressed by an order of magnitude compared with the system using traditional designs, showing a promising and practicable way to realize highly stable continuous-variable quantum key distribution system on chip. 
\end{abstract}

\setboolean{displaycopyright}{false} 

\begin{document}

\maketitle

Quantum key distribution (QKD) \cite{Bennett_BB84_1984} enables secret key generation between two distant legitimate parties with information theoretical security \cite{Pirandola_Advances_2020,Xu_RevModPhys_2020,portmann_RevModPhys_2022}. The large-scale application of QKD requires the system with high compatibility, tiny size and scaled production capacity, which can be achieved by photonic integrated circuit (PIC) techniques \cite{wang2020integrated,luo2023recent}.
Among the many QKD systems on chip, continuous-variable QKD (CV-QKD) using coherent light \cite{Grosshans_PhysRevLett_2002, Weedbrook_PhysRevLett_2004} is a promising way that provides the ability of using key devices compatible with classical optical communications \cite{zhang2023continuous}, which is favorable for the integration and large-scale application \cite{Jouguet_NatPhotonics_2013,Huang_OptLett_2015,Zhang_QuantumSciTechnol_2019,Wang_OptExpress_2020,Zhang_PhysRevLett_2020,tian2022experimental,Pi2023SubMbps,hajomer2024long}. 
Until now, the feasibility of the chip-based CV-QKD systems has been verified by an in-line local oscillator (LO) system where all of the key devices except for the laser source are integrated on the Silicon-on-Insulator (SOI) chip \cite{Zhang_NatPhotonics_2019}. Recently, the high-performance local LO CV-QKD light source \cite{li2023continuous}, transmitter \cite{aldama2023inp}, dynamic polarization controller \cite{wang2022silicon} and receiver \cite{pietri2023cv,hajomer2023continuous,bian2024continuousvariable} on chip have been brought out, which shows the great potential of realizing high-speed QKD with mature PIC process.

The stable and precise quantum state preparation is of significant importance for the chip-based CV-QKD systems \cite{Zheng_PhysRevA_2019}, where the optical power is normally less than -60 dBm at 1550 nm for the quantum signals with several photons averagely.
Any error or fluctuation will lead to extra excess noise, which limits the system performance and raises the difficulty of error correction. Moreover, it may lead to an inaccurate parameter estimation that leaves practical security loopholes.
To satisfy the requirements above, the quantum state preparation normally consists of a modulator and a variable optical attenuator (VOA), where the former is responsible for loading information and the latter is used for power control  \cite{Jouguet_NatPhotonics_2013,Huang_OptLett_2015,Zhang_QuantumSciTechnol_2019,Wang_OptExpress_2020,Zhang_PhysRevLett_2020,tian2022experimental,Pi2023SubMbps,hajomer2024long}.
In discrete systems, the VOA is typically mechanical.
Unfortunately, for the system on chip, the VOA needs to be well compatible with the manufacturing process, which is unfavorable for the integration of the high-precision mechanical structures. The current solution is to use on-chip interferometers with phase shifters to achieve the extremely high extinction ratio \cite{ma2016silicon,Zhang_NatPhotonics_2019}, however, the stability and precision are less concerned, which may be affected by the environment such as the temperature and significantly affects the consistency of the long-term system operation, performance and practical security.

In this work, a highly stable quantum signal power control for CV-QKD systems on chip is proposed based on a specially designed biased Mach-Zehnder interferometer (MZI) scheme, which is then experimentally manufactured with the standard SOI processing and tested in a CV-QKD transmitter for 80 minutes continuously. 
Theoretical analysis shows that the traditional chip-based VOA designs for QKD systems \cite{ma2016silicon,Zhang_NatPhotonics_2019} lead to an extremely sensitive and instable quantum signal power control, which may cause the limitation of system performance, and far more seriously, the overestimation of secret key rate that threaten the practical system security.
Experimentally, the proposed scheme can effectively reduce the sensitivity to the environment. The free-running experimental results show that the CV-QKD transmitter with the designed chip-based power control can achieve a stable secret key rate of 1.97E-3 bits/pulse at 60 km, of which the standard deviation is 3.77E-6 bits/pulse, an order of magnitude lower than that of the traditional chip-based VOA designs \cite{ma2016silicon,Zhang_NatPhotonics_2019}. The achievement provides a promising and practicable solution to achieve a stable quantum signal power control on photonic integrated chips, which closes the practical security loopholes and contributes to the enhancements on performance and stability simultaneously.

\begin{figure}[t]
  \centering
  \fbox{\includegraphics[width=7 cm]{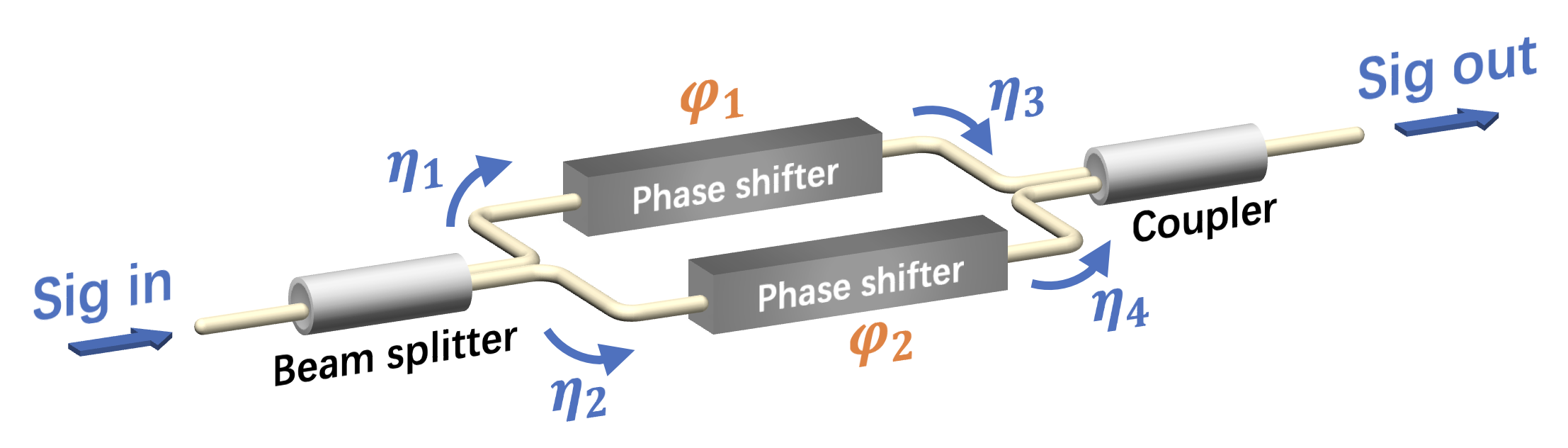}}
  \caption{The scheme of the VOA with a MZI structure. The input signal is divided by a beam splitter, where the transmittance of the two arms are $\eta_1$ and $\eta_2$. The phase shifters in the up and down arms set the phase of the output signals as $\varphi_1$ and $\varphi_2$, which are then coupled by the coupler with transmittance of $\eta_3$ and $\eta_4$.}
  \label{fig:MZIScheme}
\end{figure}
The VOA for QKD systems on chip using the most widely used MZI scheme \cite{ma2016silicon,Zhang_NatPhotonics_2019} is shown in Fig. \ref{fig:MZIScheme}, when supposing the input signal as 
\begin{equation}
  E_{in}=A e^{j \omega t},
  \label{eq:Q_in}
\end{equation}
the signals after the beam splitter can be written as $E_1=\sqrt{\eta_1}A e^{j \omega t}$ and $E_2=\sqrt{\eta_2}Ae^{j\omega t}$.
After the phase shifters, the signals are $E^{'}_{1} = \sqrt{\eta_1}Ae^{j (\omega t + \varphi_1)}$ and $E^{'}_{2} = \sqrt{\eta_2}Ae^{j (\omega t + \varphi_2)}$. Therefore, the output signal after the coupler can be written as 
\begin{equation}
  \begin{aligned}
    E_{out}=\sqrt{\eta_3}E^{'}_{1}+\sqrt{\eta_4}E^{'}_{2} = Ae^{j\omega t}(\sqrt{\eta_1 \eta_3}e^{j\varphi_1}+\sqrt{\eta_2 \eta_4}e^{j\varphi_2}),
  \end{aligned}
  \label{eq:Q_out}
\end{equation}
where the power of $E_{out}$ is
\begin{equation}
  P_{out}=A^2(\eta_1 \eta_3 + \eta_2 \eta_4 + 2\sqrt{\eta_1 \eta_2 \eta_3 \eta_4} cos \Delta \varphi).
  \label{eq:P_out}
\end{equation}
Here, $\Delta \varphi = \varphi_1-\varphi_2$.
With the input optical power $P_{in} = A^2$, the loss introduced by the VOA can be achieved with $\alpha = |10log_{10}(P_{out}/P_{in})|$.
For the VOA with a symmetrical structure where $\eta_1=\eta_2=\eta_3=\eta_4=\eta_0$, we can get 
\begin{equation}
  \alpha (\Delta \varphi) =  |10log_{10}[2 \eta^2_0 (1 +  cos \Delta \varphi)]|.
  \label{eq:alpha_sym}
\end{equation}
Suppose $\eta_0=0.5$, the minimum and the maximum attenuation can be achieved with $\alpha_{min} = \alpha(0) = 0$ and $\alpha_{max} = \alpha(\pi) = \infty$, which is the maximum adjustable range that a VOA can realize. 

\begin{figure}[t]
  \centering
  \fbox{\includegraphics[width=7 cm]{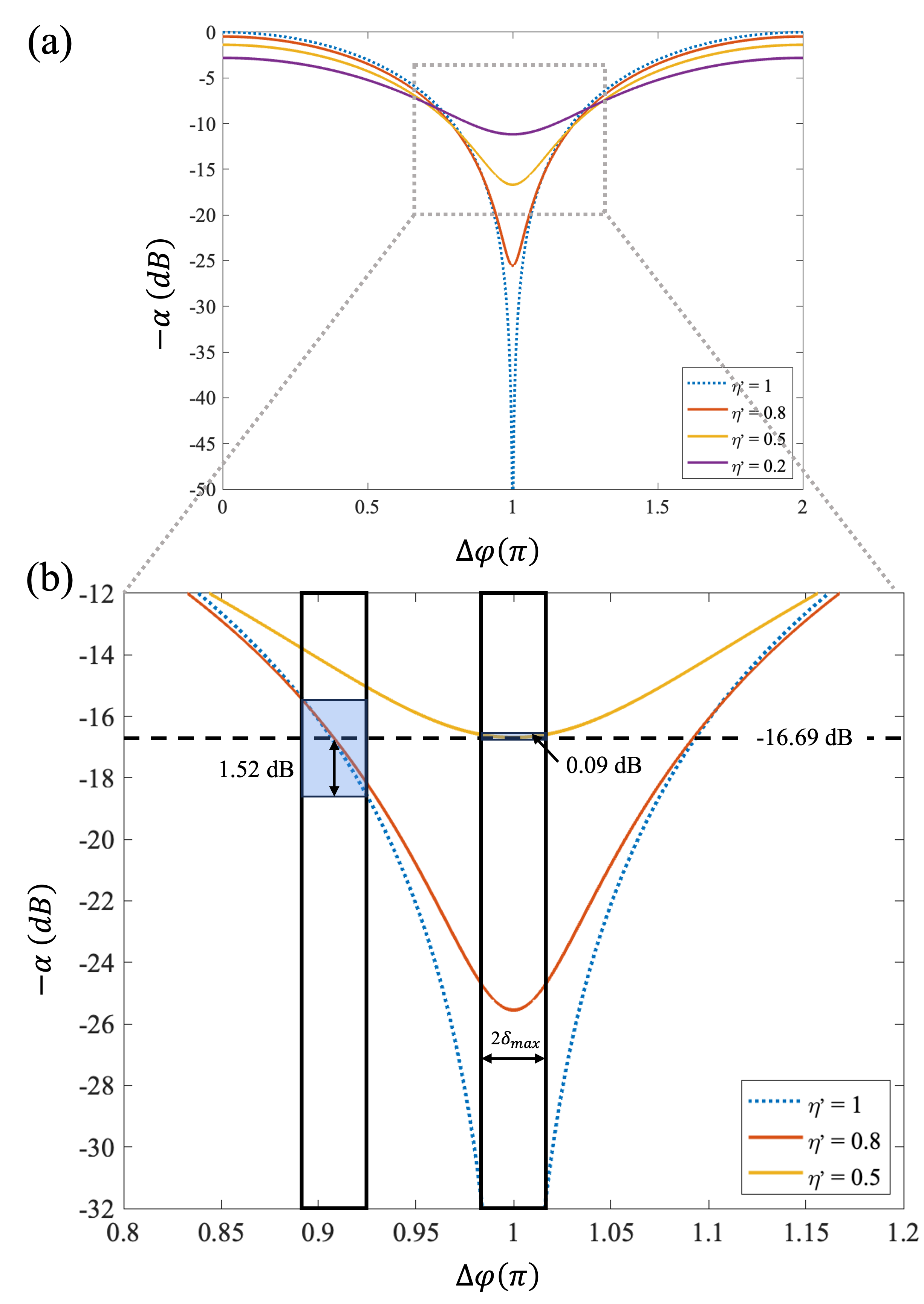}}
  \caption{The simulation results of the VOA with MZI scheme. (a) The curve of attenuation ($-\alpha$) versus phase difference ($\Delta \varphi$), in the symmetrical case ($\eta^{,} = 1$, blue dashed line) and the asymmetrical case with different $\eta^{,}$.
  (b) The fluctuation range of $\alpha$. Here, $\delta_{max} = 0.016\pi$, $\alpha_0 = 16.69$ dB, $\Delta \alpha_{max}$ with $\eta^{,} = 1$ and $\eta^{,} = 0.5$ are 1.52 dB and 0.09 dB, respectively. 
  }
  \label{fig:Simulation_VOA}
\end{figure}

However, in practical CV-QKD systems, fluctuation of $\Delta \varphi$ is inevitably introduced, denoted as $\delta$. 
It consists of two components: the slow-fading part, which is influenced by environmental temperature changes, and the fast-fading part, caused by the noise in the electrical driving signals.
In this situation, $\Delta \varphi$ will fluctuate in the range of $[\Delta \varphi - \delta_{max}, \Delta \varphi + \delta_{max}]$, leading to the fluctuation of $\alpha$. 
When denoting the target extinction ratio as $\alpha_0$, we can get $\Delta \alpha_{max}$ that represents the maximum deviation from $\alpha_0$. A smaller $\Delta \alpha_{max}$ means a narrower range of the fluctuation of $\alpha$, which results in a system with better stability.
For this purpose, the VOA should be insensitive to $\Delta \varphi$, while unfortunately, the symmetrical scheme is sensitive to the $\Delta \varphi$ as shown in Fig. \ref{fig:Simulation_VOA} (a). For $\alpha>15$ dB, the small fluctuation of $\Delta \varphi$ can lead to a large change of $\alpha$, which significantly limits the system stability.

\begin{figure*}[t]
  \centering
  \fbox{\includegraphics[width = 15 cm]{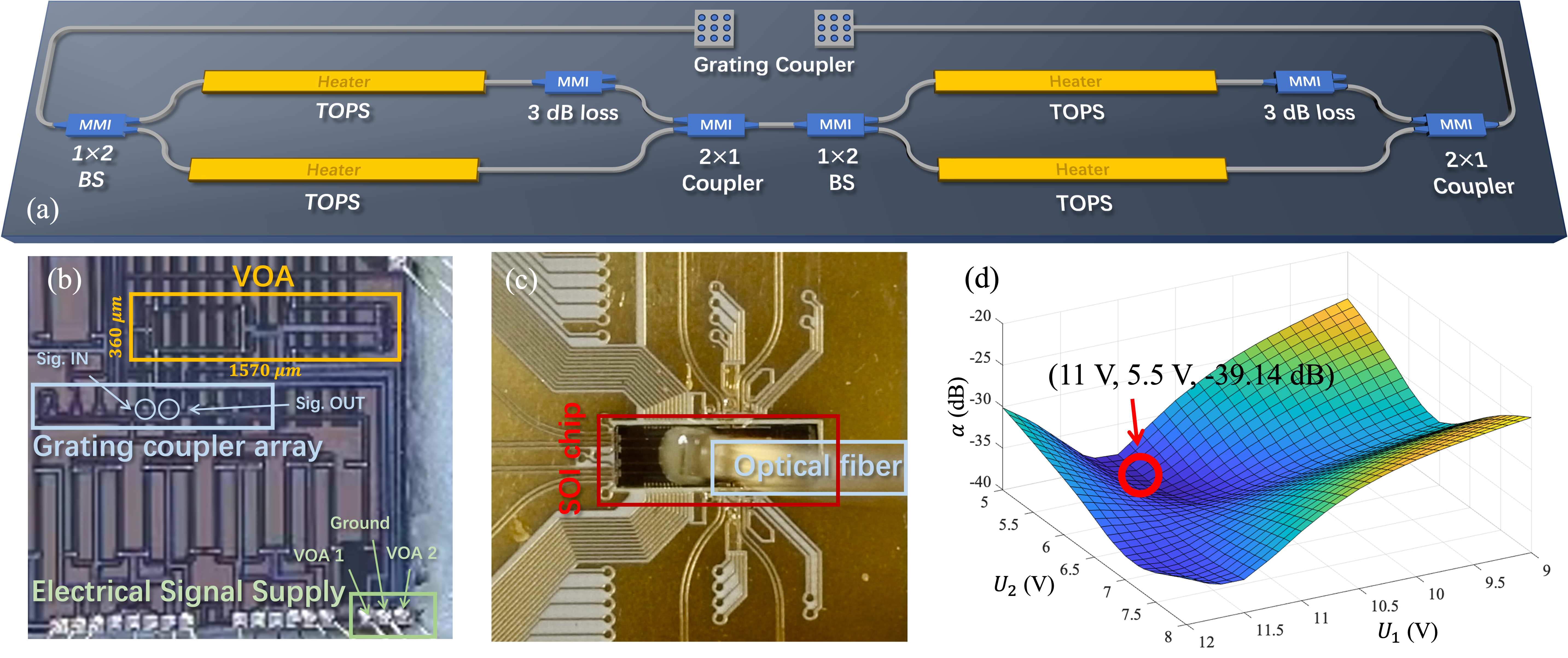}}
  \caption{The highly stable chip-based VOA for CV-QKD systems. (a) The scheme of the chip-based VOA. (b) Microscopic photo of the SOI chip. (c) The on-chip power control system after packaging. (d) $\alpha$ of the VOA corresponding to different electrical control voltages adding to the two MZIs (denoted as $U_i$). The maximum $|\alpha| = 39.14$ dB is achieved with $U_1 = 11$ V and $U_2=5.5$ V. MMI: multi-mode interferometer, TOPS: thermo-optic phase shifter, BS: beam splitter.}
  \label{fig:Scheme_VOA}
\end{figure*}

To realize a highly stable VOA on chip, the VOA with a biased MZI scheme is proposed, which introduces an extra loss on one arm of the interferometer, where the equivalent transmittance can be denoted as $\eta^{,}$.  
Without loss of generality, suppose $\eta_3 = \eta_0 \eta^{,}$, Eq. \ref{eq:alpha_sym} can be corrected as 
\begin{equation}
  \alpha(\Delta \varphi) = |10log_{10} [\eta^2_0(1+\eta^{,}+2 \sqrt{\eta^{,}} cos \Delta \varphi)]|.
  \label{eq:alpha2}
\end{equation}
A higher level of asymmetry means a lower $\eta^{,}$, which results in a less sensitive response to $\Delta \varphi$ as shown in Fig. \ref{fig:Simulation_VOA} (a). 
As opposed to the traditional design, the VOA with a biased MZI has an insensitive response to $\Delta \varphi$ at the maximum attenuation point.
When $\alpha_0$ and fluctuation range ($2\delta_{max}$) are same, as shown in Fig. \ref{fig:Simulation_VOA} (b), $\Delta \alpha_{max}$ between the VOA with biased and symmetrical schemes has a significant difference, where the biased one can suppress the fluctuation of $\alpha$ by 2 orders of magnitude.
The cost is that the adjustable range of the unbalanced VOA is limited, which should be connected to realize high $\alpha_{max}$.




Based on the above theoretical analysis, a VOA on chip for stable power control in a CV-QKD system is manufactured using the 180 nm process technology node of the standard 220 nm SOI platform \cite{SITRI}. The scheme is shown in Fig. \ref{fig:Scheme_VOA} (a), where the whole VOA consists of two biased MZI structures and each is designed to realize at least 15 dB attenuation. The theoretical analysis above suggests that $\eta^{,}=0.5$ is a suitable bias coefficient which achieves the balance between the attenuation range and the fluctuation. It is also favorable for the practical implementation on chip, where the additional transmittance of 0.5 is realized with a multi-mode interferometer (MMI) on one of the two arms. To achieve high precision, a heater with the resistance of 3 k$\Omega$ is used. Using a direct-current (DC) power supply of which the accuracy is 1 mV, we can realize the stable control of the optical power with the precision of 0.01 dBm.

\begin{figure}[b]
  \centering
  \fbox{\includegraphics[width = 7 cm]{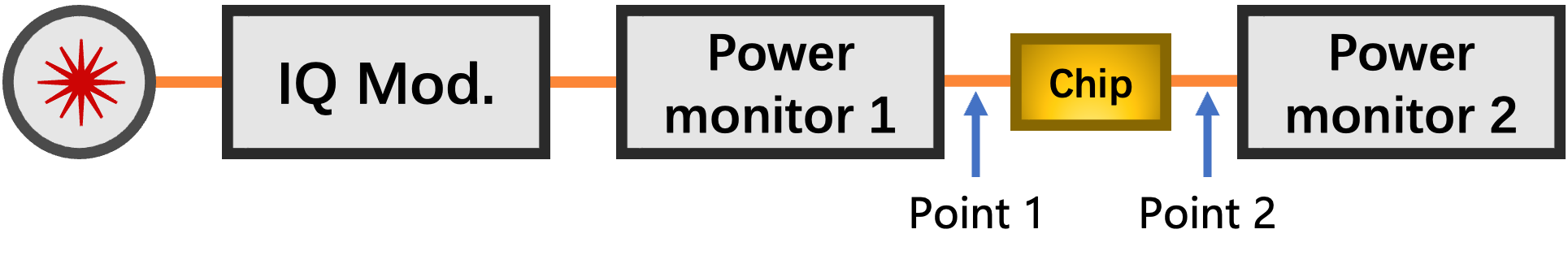}}
  \caption{The test scheme of the chip-based VOA. The coherent light from the laser source is firstly Gaussian modulated with the IQ modulator, which is then sent into the VOA chip. The optical powers of point 1 and point 2 are monitored.}
  \label{fig:Scheme_System}
\end{figure}

The input and output of the optical signal on chip is supported by a grating coupler array as shown in Figs. \ref{fig:Scheme_VOA} (b) and (c), where the loss of each grating coupler is 4 dB theoretically. 
The biased structure limits the minimum adjustable extinction ratio, which is 1.38 dB.
Thus, for the whole VOA chip, the designed adjustment range of the attenuation is 9.38 dB to 38 dB. Fig. \ref{fig:Scheme_VOA} (d) which presents the attenuation performance near the maximum attenuation point shows the well fit between the theory and experiment, where the maximum $|\alpha|$ can reach 39.14 dB with proper electrical control signals. The minimum attenuation can be realized with the electrical voltages of 6.7 and 11.1 V for the first and the second biased MZI, and the experimental result is 12.73 dB.
Note that, the extra loss is caused by the fiber connection, the imbalance of the MMI, and the additional insertion loss of the
chip-based devices.

\begin{figure}[t]
  \centering
  \fbox{\includegraphics[width= 8 cm]{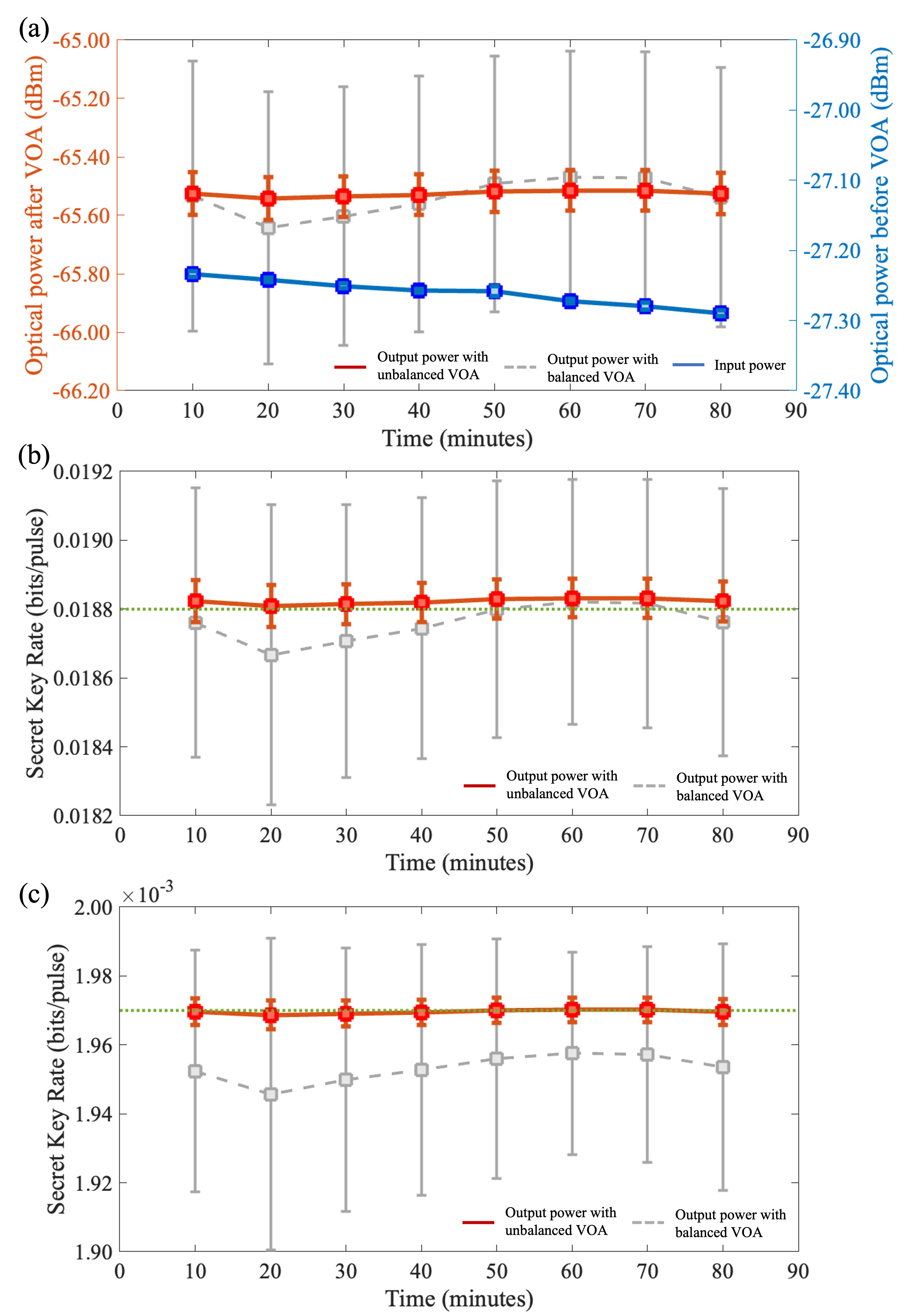}}
  \caption{The experimental and simulation results. (a) The output and input power of the VOA chip. (b) The secret key rate at 30 km. (c) The secret key rate at 60 km. Each point is achieved with the data of 10 minutes, the total length of the test is 80 minutes. 
  The data of the unbalanced VOA is simulated with the experimental data measured in the tests of unbalanced VOA.
  The secret key rate is simulated with data length of $10^9$, excess noise of 0.05 SNU, reconciliation efficiency of $95.6 \ \%$, detection efficiency of 0.6 and electronic noise of 0.1 SNU. The aiming secret key rate is the dashed green line.}
  \label{fig:Results_System}
\end{figure}

The chip-based VOA is then tested in a standard CV-QKD transmitter as shown in Fig. \ref{fig:Scheme_System}.
The repetition frequency of the quantum signal is 1 GHz, and the laser is working at 1550.12 nm. 
The modulation variance is set as 4.4 shot noise unit (SNU) for the optimization of the system, corresponding to the optical power of point 2 of -65.50 dBm. The mean value of the input power of the VOA chip is controlled around -27.26 dBm, therefore, the required attenuation is around 38 dB, which can be covered by the designed VOA. In this way, each unbalanced MZI is working near the maximum attenuation point, which contributes to a stable performance. 

The input and output optical power of the unbalanced chip-based VOA is tested for 80 minutes with data collected every second, as shown in Fig. \ref{fig:Results_System} (a). 
We remark that the tested chip-based VOA is working at the free-running mode without any feedback control.
To make a comparison, the extinction ratio of the VOA with symmetrical MZI is simulated with the experimental phase derivation $\delta$ of every second. It is calculated with the experimental data using Eq. \ref{eq:alpha2}, supposing $\Delta \varphi = \Delta \varphi_0 +\delta$, where $\alpha_0 = \alpha(\Delta \varphi_0)$. For the simulated VOA with symmetrical MZI, the target extinction ratio is set as 30 dB, which is affected by the $\delta$ of every second calculated above. In this way, the simulation condition is set as the same of the experiment, which contributes to a fair comparison.
Specifically, the mean value and the standard deviation of the VOA with biased MZI is 65.52 dB and 0.071 dB, while 65.54 dB and 0.45 dB for the simulated VOA with symmetrical MZI, of which the standard deviation is about an order of magnitude higher.

Figs. \ref{fig:Results_System} (b) and (c) show the fluctuation of secret key rate when the output power of the transmitter is fading caused by the imperfect VOA. The secret key rate is calculated considering the finite-size effect with one-time SNU calibration \cite{Zhang_QuantumSciTechnol_2019,Zhang_PhysRevApplied_2020},
\begin{equation}
  K = \beta I_{AB}-\chi_{BE}-\Delta(n).
\end{equation}
Here, $\beta$ is the reconciliation efficiency, $I_{AB}$ is the classical mutual information between the transmitter and receiver, $\chi_{BE}$ is the upper bound of Eve's knowledge on Bob and $\Delta(n)$ is related to the security of the privacy amplification. 
The transmission distance is set as 30 km and 60 km, and the system parameter is set as the typical values achieved in system experiments \cite{Zhang_PhysRevLett_2020}. The ideal secret key rates at 30 km and 60 km are 1.88E-2 bits/pulse and 1.97E-3 bits/pulse with a modulation variance of 4.4 SNU. 
In this experiment, at 30 km, the mean values of the secret key rates when using VOAs with biased and symmetrical MZI are both 1.88E-2 bits/pulse, which fit well with the target value. However, the standard deviation of the above two schemes are 5.8E-5 bits/pulse and 3.9E-4 bits/pulse, respectively. At 60 km, the mean values of the secret key rates are 1.97E-3 bits/pulse and 3.95E-3 bits/pulse, where the VOA with biased MZI still fits the target secret key rate but the balanced VOA leads to a deviation. This results in an overestimation of more than 1 \% when the parameter estimation is performed with the modulation variance with a set value of 4.4 SNU.
The standard deviation of the secret key rates are 3.77E-6 and 3.62E-5, which is 0.19 \% and 1.84 \% of the target value. In conclusion, the proposed VOA with biased MZI contributes to a stable experimental system performance where the standard deviation of secret key rate is about an order of magnitude lower than that of the traditional design.

In this work, we have closed the practical security loopholes caused by the unstable on-chip power control for CV-QKD systems. The experimental and simulation results have shown the significant enhancements of the system stability and performance. Remark that, the photonic chip is continuously tested in a free-running manner without any feedback control and temperature control, which reflects the high stability of the designed structure itself. Thermoelectric coolers and source monitor can be used to further stabilize the system, and it is expected to stabilize the secret key rate with standard deviation lower than 0.1 \%. The chip-based quantum signal power control can also be used at the detector site for the balance control of homodyne detection, which is of significant importance for a low-noise measurement. It is expected that the designed chip-based VOA will effectively improve the system stability and overall performance, which contributes to a stable practical chip-based CV-QKD system.


\begin{backmatter}
\bmsection{Funding} The National Key Research and Development Program of China (Grant No. 2020YFA0309704), the National Natural Science Foundation of China (Grant Nos U22A2089, 62101516, 62171418, 62201530 and 61901425), the Sichuan Science and Technology Program (Grant Nos 2023JDRC0017, 2023YFG0143, 2022ZDZX0009 and 2021YJ0313), the Natural Science Foundation of Sichuan Province (Grant Nos 2023NSFSC1387 and 2023NSFSC0449), the Basic Research Program of China(Grant No. JCKY2021210B059), the Equipment Advance Research Field Foundation(Grant No. 315067206), the Chengdu Key Research and Development Support Program (Grant No 2021-YF09-00116-GX).


\bmsection{Disclosures} The authors declare no conflicts of interest.

\bmsection{Data availability} Data underlying the results presented in this paper are not publicly available at this time but may be obtained from the authors upon reasonable request.


\end{backmatter}

\bibliography{sample}

\bibliographyfullrefs{sample}


\ifthenelse{\equal{\journalref}{aop}}{%
\section*{Author Biographies}
\begingroup
\setlength\intextsep{0pt}
\begin{minipage}[t][6.3cm][t]{1.0\textwidth} 
  \begin{wrapfigure}{L}{0.25\textwidth}
    \includegraphics[width=0.25\textwidth]{john_smith.eps}
  \end{wrapfigure}
  \noindent
  {\bfseries John Smith} received his BSc (Mathematics) in 2000 from The University of Maryland. His research interests include lasers and optics.
\end{minipage}
\begin{minipage}{1.0\textwidth}
  \begin{wrapfigure}{L}{0.25\textwidth}
    \includegraphics[width=0.25\textwidth]{alice_smith.eps}
  \end{wrapfigure}
  \noindent
  {\bfseries Alice Smith} also received her BSc (Mathematics) in 2000 from The University of Maryland. Her research interests also include lasers and optics.
\end{minipage}
\endgroup
}{}

\end{document}